\documentclass[sigconf,nonacm]{acmart}

\AtBeginDocument{%
  \providecommand\BibTeX{{%
    \normalfont B\kern-0.5em{\scshape i\kern-0.25em b}\kern-0.8em\TeX}}}

\usepackage{lipsum}
\usepackage{subcaption}
\usepackage{macros}
\usepackage{float}
\usepackage{wrapfig}
\usepackage{enumitem}

\begin{document}

\title{Diffusion User Interfaces}
\title{Gradually Generating User Interfaces as a Design Method for }
\title{Gradually Generating User Interfaces: A Design Method for Understanding Interfaces}
\title{Gradually Generating User Interfaces: A Design Method for Task-Specific Understanding of UIs}
\title{Gradually Generating User Interfaces as a Design Method for End-User Customization}
\title{Gradual Generation of User Interfaces as a Design Method for End-User Customization}
\title{Gradual Generation of User Interfaces as a Design Method for AI-Driven Malleable Software}
\title[Gradual Generation of User Interfaces]{Gradual Generation of User Interfaces as a Design Method for Malleable Software}

\author{Bryan Min}
\email{bdmin@ucsd.edu}
\affiliation{%
  \institution{University of California San Diego}
  \streetaddress{9500 Gilman Dr}
  \city{La Jolla}
  \state{California}
  \country{USA}
  \postcode{92093}
}

\author{Peiling Jiang}
\email{peiling@ucsd.edu}
\affiliation{%
  \institution{University of California San Diego}
  \streetaddress{9500 Gilman Dr}
  \city{La Jolla}
  \state{California}
  \country{USA}
  \postcode{92093}
}

\author{Zhicheng Huang}
\email{zhichenghuang@ucsd.edu}
\affiliation{%
  \institution{University of California San Diego}
  \streetaddress{9500 Gilman Dr}
  \city{La Jolla}
  \state{California}
  \country{USA}
  \postcode{92093}
}

\author{Haijun Xia}
\email{haijunxia@ucsd.edu}
\affiliation{%
  \institution{University of California San Diego}
  \streetaddress{9500 Gilman Dr}
  \city{La Jolla}
  \state{California}
  \country{USA}
  \postcode{92093}
}

\renewcommand{\shortauthors}{Min et al.}

\begin{abstract}
AI is growing increasingly capable of automatically generating user interfaces (GenUI) from user prompts. However, designing GenUI applications that enable users to discover diverse customizations while preserving GenUI's expressiveness remains challenging. Current design methods---presenting prompt boxes and leveraging context---lack affordances for customization discovery, while traditional menu-based approaches become overly complex given GenUI's vast customization space. We propose Gradually Generating User Interfaces---a design method that structures customizations into intermediate UI layers that AI gradually loads during interface generation. These intermediate stages expose different customization features along specific dimensions, making them discoverable to users. Users can wind back the generation process to access customizations. We demonstrate this approach through three prototype websites, showing how designers can support GenUI's expanded customization capabilities while maintaining visual simplicity and discoverability. Our work offers a practical method for integrating customization features into GenUI applications, contributing an approach to designing malleable software.

\end{abstract}

%


%

\begin{teaserfigure}
    \includegraphics[width=\textwidth]{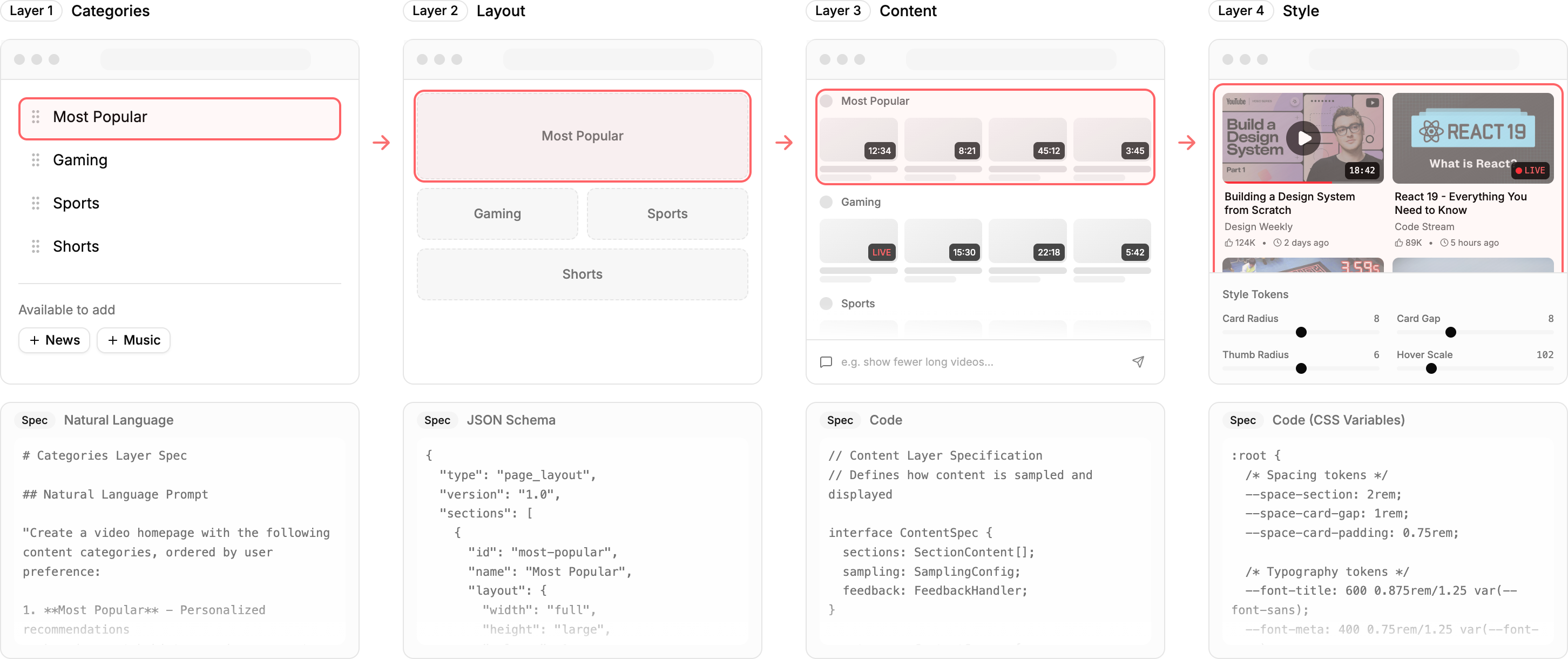}
    \caption{We present Gradual Generation as a method for designing customizations in generative UI applications. Gradual Generation proposes UI/UX designers to organize UI customizations into multiple intermediate layers that AI progressively loads during interface generation, enabling users to discover and access customizations by rewinding to earlier stages.}
    \label{fig:teaser}
    \Description{Four side-by-side panels illustrate ``Gradual Generation'' for customizing a video feed homepage through intermediate layers, with arrows showing a left-to-right progression. Layer 1 (Categories) shows a list of feed categories (e.g., ``Most Popular,'' ``Gaming,'' ``Sports,'' ``Shorts'') with controls to reorder and add new categories. Layer 2 (Layout) shows those categories rearranged and resized into sections, emphasizing ``Most Popular.'' Layer 3 (Content) shows each section populated with video cards and controls to refine what is shown. Layer 4 (Style) shows the same feed with visual styling controls (e.g., card radius, spacing, hover scale). Beneath each layer is a corresponding specification format: natural language (Categories), JSON schema (Layout), code (Content), and CSS variables (Style).}
\end{teaserfigure}

\maketitle{}

\section{Introduction}
\label{sec:introduction}

Advances in generative AI are driving a resurgence for Generative UIs (GenUIs)---user interfaces that AI can automatically generate based on user prompts and context. In fact, we are already seeing new capabilities emerge from GenUI, such as generating custom browser tabs and websites \cite{lovable2026, gentabs2026, cao2025jelly}, generating UI widgets within chatbots \cite{openai2025chatgptapps, claude2024artifacts}, and customizing UIs from user prompts \cite{malleableODI, meridian}. This demonstrates a promising future for making software more \textit{malleable} \cite{malleableODI, litt2025malleable} and easily customizable according to users' needs. 

However, it remains unclear how to best design GenUI so that users can discover diverse customizations while preserving GenUI's expressiveness. Current design presents prompt boxes and leverages background context to automatically generate UIs. This paradigm lacks the affordances necessary for users to discover customization options~\cite{hari2024gulfenvisioning, zamfirescu2023johnnyprompt}. On the other hand, if designers revert to traditional UI customization methods by organizing sets of options into menu categories, the interface becomes overly complex given the vast space of possible GenUI customizations \cite{mackay1991triggers}. This challenge calls for the HCI community to explore new paradigms for integrating customization features into GenUI applications.

We argue that the transition from prompt to generated UI itself presents an opportunity to surface diverse customization options. We posit that providing intermediate stages between the user's initial prompt and the final generated interface creates a valuable design space for organizing different categories of UI customizations.

In this paper, we aim to explore \textbf{Gradually Generating User Interfaces}---a design method that structures sets of customizations into intermediate UI ``layers'' that AI gradually loads during the process of generating the final interface.
These intermediate UIs expose different sets of customization features along specific dimensions, making them discoverable to users. Users can then ``wind back'' the gradual generation process to access the customizations they need. This approach enables designers to support the broader space of customizations that GenUI promises while maintaining both visual simplicity and discoverability (Fig. \ref{fig:teaser}).

We first introduce the Gradual Generation design method, then demonstrate its potential through three prototype websites. Our prototypes show how UI/UX designers can effectively integrate GenUI's expanded customization capabilities without sacrificing interface usability. Finally, we discuss design implications and future work for this approach to achieving malleable software with AI.

\section{Gradual Generation of User Interfaces}

Gradual Generation is a GenUI design approach that enables users to discover possible customizations in an interface by presenting them while the UI generates step-by-step.
Upon entering a GenUI page, the interface presents intermediate UIs that show key classes of customizations. For example, a video homepage illustrated in Fig. \ref{fig:teaser} may present four intermediate UIs before presenting the final interface: the user's preferred video \textit{categories}, the \textit{layout} of these categories, the \textit{content} to show in each video, and finally the \textit{style}.
The \textit{categories} layer can allow users to modify the ranking of their preferences, the \textit{layout} layer can allow users to resize and rearrange the categories on the page, the \textit{content} layer can allow users to determine which attributes are shown or hidden in each video, and the \textit{style} layer can allow users to personalize the aesthetic of their page.
This design method differs from the traditional one in that designers are responsible for identifying the key intermediate UIs, as opposed to categorizing the set of customizations.


\begin{figure}
    \centering
    \includegraphics[width=\linewidth]{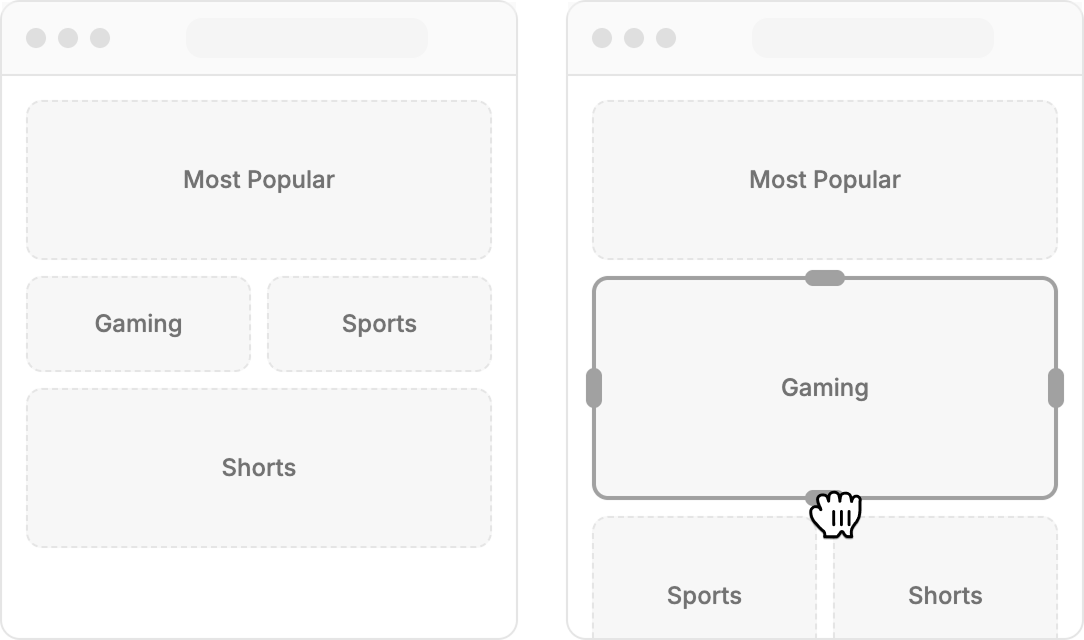}
    \caption{Intermediate UI layers provide specific sets of customizations. For example, designers can dedicate a ``Layout'' layer to customizations that let users resize, rearrange, and combine sections in the UI.}
    \label{fig:intermediate-uis}
\end{figure}







\begin{figure*}
    \centering
    \includegraphics[width=\linewidth]{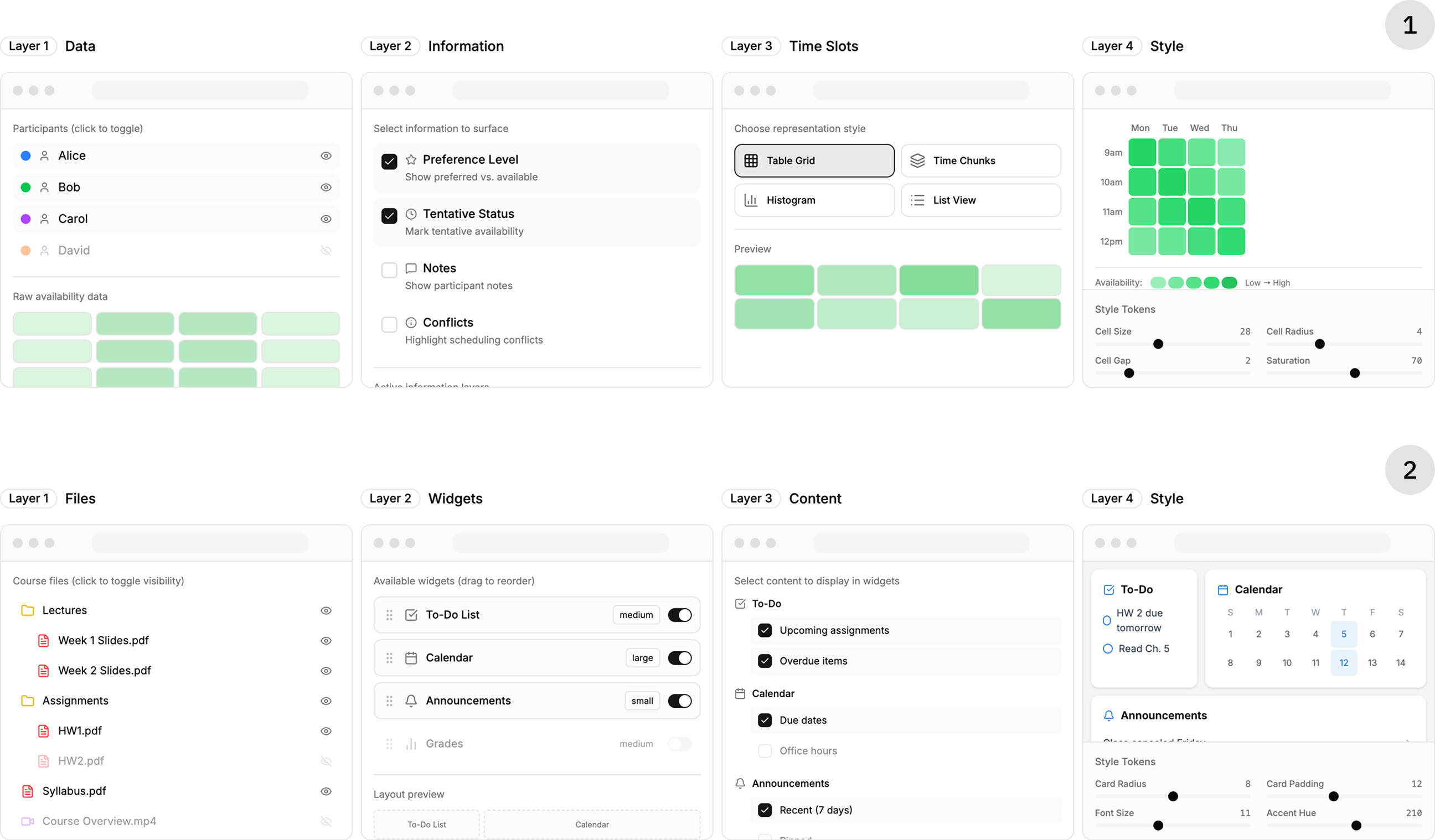}
    \caption{Example intermediate layers of (1) a team scheduling calender interface, and (2) course management system interface.}
    \Description{Example intermediate layers of (1) a team scheduling calender interface, and (2) course management system interface.}
    \label{fig:demos}
\end{figure*}

\subsection{The Gradual Generation Design Method}

The Gradual Generation design method involves four main tasks for designers. To illustrate these tasks, we continue the example of designing a video homepage UI presented in Fig. \ref{fig:teaser}.

\paragraph{\textbf{(1) Identify key intermediate stages of the UI}}

Designers first identify key intermediate stages of the UI by reflecting on their design process for producing UI mockups from raw data. For example, when designing a video homepage UI, intermediate stages might include: a categories stage that lists category rankings from the software's recommendation algorithm, followed by stages for visual layout of individual categories, then video content and data attributes within each category, and finally aesthetic styling of the homepage.

\paragraph{\textbf{(2) Design customizations for each intermediate stage}}

For each stage, designers explore and design customizations specific to that level of the interface (Fig. \ref{fig:intermediate-uis}). For instance, the category rankings stage might support customizing ranking order, introducing additional categories, or refining categories into subcategories (e.g., specifying "gaming" as "e-sports").

\paragraph{\textbf{(3) Develop specifications for customizations in each stage}}

Designers then create specifications documenting the available customizations for each intermediate stage. These specifications can take various forms: documentation pages listing all categories the application's recommender system can output, JSON type definitions, or code blocks. GenUI uses these specifications to generate interfaces based on each stage's parameter set, where UI changes map directly to specification changes.

\paragraph{\textbf{(4) Ensure smooth transitions between stages}}

Finally, designers must ensure natural progression across intermediate stages by identifying key UI elements that persist throughout the generation process. As illustrated in Fig. \ref{fig:teaser}, an individual category block from the ``Categories'' layer (Layer 1) transitions across subsequent layers, where it becomes a layout block (Layer 2), then a video section (Layer 3), and finally a component containing video thumbnails and attributes for that category (Layer 4).

\subsection{Demonstrations of Gradual Generation}



We present three demonstrative GenUI applications to show how different types of interfaces may utilize intermediate UI layers for their customization needs.

\textbf{\emph{Generative Video Feed Homepage}} benefits from intermediate layers that let users customize what they see and how it looks based on their interests (Fig.~\ref{fig:teaser}). The \textit{Categories} layer shows preferred video categories (e.g., Most Popular, Gaming, Sports) and lets users reorder, remove, or add categories. The \textit{Layout} layer lets users resize and rearrange category sections---adjusting prominence and placing sections side by side. The \textit{Content} layer controls what each section shows (e.g., filter by duration or recency) and supports natural-language feedback to refine recommendations. Finally, the \textit{Style} layer lets users personalize visual details such as card radius, thumbnail treatment, spacing, and hover effects.

\textbf{\textit{Generative Team Scheduling Calendar}} interfaces (akin to When2Meet\footnote{https://www.when2meet.com/}) can benefit from intermediate layers by allowing teams to tailor how availability is surfaced and compared (Fig.~\ref{fig:teaser}.1). The \textit{Data} layer lets users choose which participants' availability to display. The \textit{Information} layer adds optional metadata, such as granular preference levels (preferred vs.\ merely available), tentative status, or scheduling conflicts. The \textit{Time Slots} layer controls how availability is represented---as a grid, chunked time blocks, a histogram of overlap, or a list view. Finally, the \textit{Style} layer lets users adjust visual details such as cell size, border radius, and color saturation.

\textbf{\textit{Generative Course Management System}} (akin to Canvas) benefits from intermediate layers by allowing students to customize how they navigate course resources for learning (Fig.~\ref{fig:demos}.2). The \textit{Files} layer surfaces course resources and lets users show or hide specific files and folders. The \textit{Widgets} layer lets users choose which dashboard components to display (e.g., To-Do lists, calendars, announcements, grade summaries) and adjust their size and arrangement. The \textit{Content} layer controls what appears within each widget, such as showing only upcoming assignments or filtering announcements to the past week. Finally, the \textit{Style} layer lets users customize visual details such as card styling, typography, and accent colors.

\section{Discussion}

\subsection{Reconsidering the Role of UI/UX Designers in Gradually Generated UIs}

Gradual Generation reconsiders the role of UI/UX designers in GenUI systems.
Rather than focusing solely on specifying fixed user journeys or static interface mockups, designers may increasingly be responsible for structuring how interfaces can be generated, adapted, and customized, gradually.

In the context of Gradually Generated UIs, the role of UI/UX designers shifts toward defining intermediate layers that shape the space of expressiveness available to GenUI. By designing such intermediate layers, designers can articulate which aspects of an interface are open to variation, which are constrained, and along what dimensions customizations may occur.

As a result, GenUI becomes expressive through a set of layered design decisions embedded in intermediate UIs, instead of unconstrained generation.



\subsection{Enabling End-users to Add Their Own Layers in Gradual Generation}
Allowing end-users to introduce personal intermediate layers into Gradual Generation may support deeper personalization and long-term adaptation beyond what UI/UX designers can anticipate.
Such personal layers can be designed as pluggable components that integrate into the Gradual Generation process without requiring changes to the underlying system or designer-defined layers. By supporting integrating intermediate layers defined by both designers and end-users, the Gradual Generation process promotes both parties to contribute to the negotiation of control and expressiveness within the interface.

Admittedly, allowing end-users to bring in personal intermediate layers may undermine the design control intended by UI/UX designers. To strike a balance, UI/UX designers can encode a higher-order layer that selectively applies end-users' personal layers to the Gradual Generation process.

Personal layers can be derived from the gradual use of Gradually Generated User Interfaces. The customizations end-users make within intermediate layers for a given website may be valuable beyond this site. Over time, these editing traces may reflect recurring preferences and workflows. Gradual Generation allows such traces to be abstracted into personal layers that externalize users' preferences across sessions and contexts.

\subsection{Cross-platform Scalability with Re-usable Platform-specific Intermediate Layers}
Florins and Vanderdonckt~\cite{florins2004graceful} proposed a design method where a user interface designed for one platform can be adapted for another platform by applying transformational rules.
We argue that Gradual Generation is scalable by incorporating re-usable platform-specific intermediate layers into existing Gradually Generated UIs to apply to other use cases.
For example, in order to re-use a Gradually Generated UI to a new platform that supports a unique haptics modality, the UI/UX designer can define boundaries of constraints and degree of customization for that specific haptic modality in a platform-specific intermediate layer to be integrated with the existing layers, while preserving the rest of the design system.



\section{Future Work}
\label{section:conclusion}

Our goal is to contribute design guidelines for UI/UX designers to effectively design GenUI into their applications, making customizations more discoverable. To further investigate the opportunity of Gradual Generation, our future work will involve answering the following research questions:

\begin{enumerate}[label=\textbf{[RQ\arabic*]}]
    \item What kinds of intermediate UI layers do UI/UX designers build for their software applications to be customizable?
    \item How might UI/UX designer workflows change when designing Gradually Generated UIs as opposed to traditional UIs?
    \item How do end-users' customizing behaviors change when given software applications that implement Gradually Generated UIs?
\end{enumerate}

\paragraph{RQ1}

We have demonstrated the feasibility of designing customizations into GenUI applications by structuring them within intermediate layers. In our current prototypes, we propose that UI/UX designers must identify and build intermediate UI layers that provide various customizations within the application. While designers are capable of identifying relevant layers, we anticipate large variability in implementation across applications. For example, customization behaviors for a "Layout" layer in YouTube may drastically differ from a "Layout" layer in Vimeo. 
In our future work, we aim to conduct a workshop that invites UI/UX designers to design intermediate UIs for their applications. We then aim to synthesize the intermediate layers to synthesize a common set and recommend best practices for future designers.

\paragraph{RQ2}

Another open question we plan to investigate involves observing how organizational workflows in design teams may shift to best support Gradually Generated UIs. We aim to observe teams of industry designers as they design Gradually Generated UIs, documenting their collaboration patterns, the UI mockups they create, and potential challenges and frictions they experience during the process. Insights from these observations may inform new UI design tools to support designers in creating applications with GenUI integrations.

\paragraph{RQ3}

The ultimate purpose of providing the Gradual Generation design method to designers is to enable more malleable UIs and observe how customization behaviors change when users gain enhanced interface customization capabilities. To study this, we aim to develop two example web applications designed using the Gradual Generation method, deploy them for end-users over a two-week period, and observe their customization patterns and navigation behaviors.

\section{Conclusion}

We introduced Gradual Generation of UIs as a design method for integrating customization features into GenUI systems. We demonstrated examples showing the potential for intermediate UIs to effectively organize sets of customizations, discussed design implications of Gradual Generation, and described future steps for investigating this design method. We hope our work encourages future research to revisit traditional methods for designing UI customizations and enables UI/UX designers to create more malleable software.




\bibliographystyle{ACM-Reference-Format}
\bibliography{main}

\appendix


\end{document}